\documentstyle[12pt,twoside,dina4,epsf,epsfig]{article}

\begin{document}

\renewcommand{\topfraction}{0.9}
\renewcommand{\bottomfraction}{0.9}
\renewcommand{\textfraction}{0.1}
\footheight0cm
\headheight0cm
\headsep1cm
\footskip1cm
\parindent0em
\textheight23cm
\newcommand{\Theli}{\mbox{T$_{\scriptsize\rm HeLi}$}}
\newcommand{\Zbound}{\mbox{Z$_{\scriptsize\rm bound}$}}
\newcommand{\lif}{\mbox{$^{5}$Li}}
\newcommand{\lis}{\mbox{$^{6}$Li}}
\newcommand{\bea}{\mbox{$^{8}$Be}}
\newcommand{\hef}{\mbox{$^{4}$He}}
\newcommand{\pp}{\mbox{$p$--$p$}}
\newcommand{\ptri}{\mbox{$p$--$t$}}
\newcommand{\palf}{\mbox{$p$--$\alpha$}}
\newcommand{\plisi}{\mbox{$p$--$^{7}$Li}}
\newcommand{\dd}{\mbox{$d$--$d$}}
\newcommand{\dalf}{\mbox{$d$--$\alpha$}}
\newcommand{\dhed}{\mbox{$d$--$^3$He}}
\newcommand{\tritri}{\mbox{$t$--$t$}}
\newcommand{\trihed}{\mbox{$t$--$^3$He}}
\newcommand{\alfalf}{\mbox{$\alpha$--$\alpha$}}

\newenvironment{ownitem}{\begin{list}{\labelitemi}{
\leftmargin0pt \listparindent2ex \itemindent2ex }}{\end{list}}
\newcounter{listenum}

\newenvironment{ownenum}{\begin{list}{\arabic{listenum}.}{
\usecounter{listenum} 
\labelwidth1ex \leftmargin0pt \listparindent2ex \itemindent3ex
}}{\end{list}}

\begin{center}
{\Large\bf Emission temperatures and freeze out densities\\[0.3ex] 
from light particle correlation functions\\[0.5ex] 
in Au+Au collisions at 1~A$\cdot$GeV}\\[2.5ex]
S.Fritz for the ALADiN collaboration:\\[1.0ex]
R.~Bassini,$^{(2)}$
M.~Begemann-Blaich,$^{(1)}$
A.S.~Botvina,$^{(3)}$\footnote{
Present Address: Bereich Theoretische Physik, Hahn-Meitner-Institut,
D-14109 Berlin, Germany}
S.J.~Gaff,$^{(4)}$
C.~Gro\ss,$^{(1)}$
G.~Imm\'{e},$^{(5)}$
I.~Iori,$^{(2)}$
U.~Kleinevo\ss,$^{(1)}$
G.J.~Kunde,$^{(4)}$
W.D.~Kunze,$^{(1)}$
U.~Lynen,$^{(1)}$
V.~Maddalena,$^{(5)}$                   
M.~Mahi,$^{(1)}$
T.~M\"ohlenkamp,$^{(6)}$
A.~Moroni,$^{(2)}$
W.F.J.~M\"uller,$^{(1)}$
C.~Nociforo,$^{(5)}$                    
B.~Ocker,$^{(7)}$
T.~Odeh,$^{(1)}$
F.~Petruzzelli,$^{(2)}$
J.~Pochodzalla,$^{(1)}$\footnote{
Present address: Max-Planck-Institut f\"ur Kernphysik,
D-69117 Heidelberg, Germany}
G.~Raciti,$^{(5)}$
G.~Riccobene,$^{(5)}$                   
F.P.~Romano,$^{(5)}$                    
Th.~Rubehn,$^{(1)}$\footnote{
Present address: Nuclear Science Division, Lawrence Berkeley Laboratory,
Berkeley, CA 94720, USA}
A.~Saija,$^{(5)}$                       
M.~Schnittker,$^{(1)}$
A.~Sch\"uttauf,$^{(7)}$
C.~Schwarz,$^{(1)}$
W.~Seidel,$^{(6)}$
V.~Serfling,$^{(1)}$
C.~Sfienti,$^{(5)}$                     
W.~Trautmann,$^{(1)}$
A.~Trzcinski,$^{(8)}$
G.~Verde,$^{(5)}$
A.~W\"orner,$^{(1)}$
Hongfei~Xi,$^{(1)}$\footnote{
Present address: NSCL, Michigan State University,
East Lansing, MI 48824, USA }
and B.~Zwieglinski$^{(8)}$\\[2.0ex]
$^{(1)}$Gesellschaft  f\"ur  Schwerionenforschung, D-64291 Darmstadt,
Germany\\
$^{(2)}$Istituto di Scienze Fisiche dell' Universit\`{a}
and I.N.F.N., I-20133 Milano, Italy\\
$^{(3)}$Institute for Nuclear Research,
Russian Academy of Sciences, 117312 Moscow, Russia\\
$^{(4)}$Department of Physics and
Astronomy and National Superconducting Cyclotron Laboratory,
Michigan State University, East Lansing, MI 48824, USA\\
$^{(5)}$Dipartimento di Fisica dell' Universit\`{a}
and I.N.F.N.,
I-95129 Catania, Italy\\
$^{(6)}$Forschungszentrum Rossendorf, D-01314 Dresden, Germany\\
$^{(7)}$Institut f\"ur Kernphysik,
Universit\"at Frankfurt, D-60486 Frankfurt, Germany\\
$^{(8)}$Soltan Institute for Nuclear Studies,
00-681 Warsaw, Hoza 69, Poland
\end{center}
\pagenumbering{arabic}
\begin{abstract}
A study of emission temperatures extracted from excited state populations
and of freeze out radii from light particle intensity interferometry is 
presented. Three high resolution
$\Delta E$--$E$--Hodoscopes with a total of 216 detectors are combined
with the ALADiN setup in order to study Au+Au collisions at 1 A$\cdot$GeV. In 
contrast to  measurements with the isotope thermometer
the extracted apparent temperatures do not vary with impact parameter thus with 
excitation energy. From the extracted  radii a freeze out density was 
determined which decreases from $0.2\rho_0$ for the most peripheral to less than
$0.1\rho_0$ for the most central collisions. A density--dependent feeding 
correction is applied to the different temperature measurements.
\end{abstract}
\nopagebreak
\section{Introduction}
In a recent publication \cite{poc95} we presented the determination 
of emission
temperatures and excitation energies in Au+Au collisions in order to 
explore the liquid-gas phase transition of nuclear matter. One open
question was the influence of feeding on the isotope thermometer
\Theli. In a follow up experiment we compared the 
isotope thermometer with temperatures derived from the relative
yields of excited particle--unbound states. The idea was to 
calibrate the isotope thermometer with a temperature determined 
from the excited states rather than deriving this calibration from 
theoretical calculations.\\[0.5ex]
The following excited state thermometers were analyzed:
\begin{enumerate}
\item \lif~which decays into \dhed~ ($E^*=16.66$~MeV) and 
into \palf~ in the ground state.
\item \lis~where the apparent temperature was determined by the ratio of the 
third ($E^*=4.31$~MeV) and fifth ($E^*=5.65$~MeV) excited states to the first
($E^*=2.168$~MeV) excited state which all decay into \dalf.
\item \bea~where the ratio of the state at $E^*=17.64$~MeV   
(decays into \plisi) to the state at $E^*=3.04$~MeV (decays into \alfalf)
is used for the temperature determination.
\end{enumerate}
The paper is organized as follows:\\[0.5ex]
In section~\ref{exp} details of the experimental setup are given. 
The extracted temperatures for various event selection criteria are 
presented in section~\ref{temp}. Extracted radii and freeze
out densities from \pp~ and \dalf~ correlations are presented in 
section~\ref{density}. In 
section~\ref{feeding} the results are compared to feeding  calculations.
\section{Experimental setup and data analysis}\label{exp}
\begin{figure}[htb]
\begin{center}
\fbox{\epsffile[40 530 400 760]{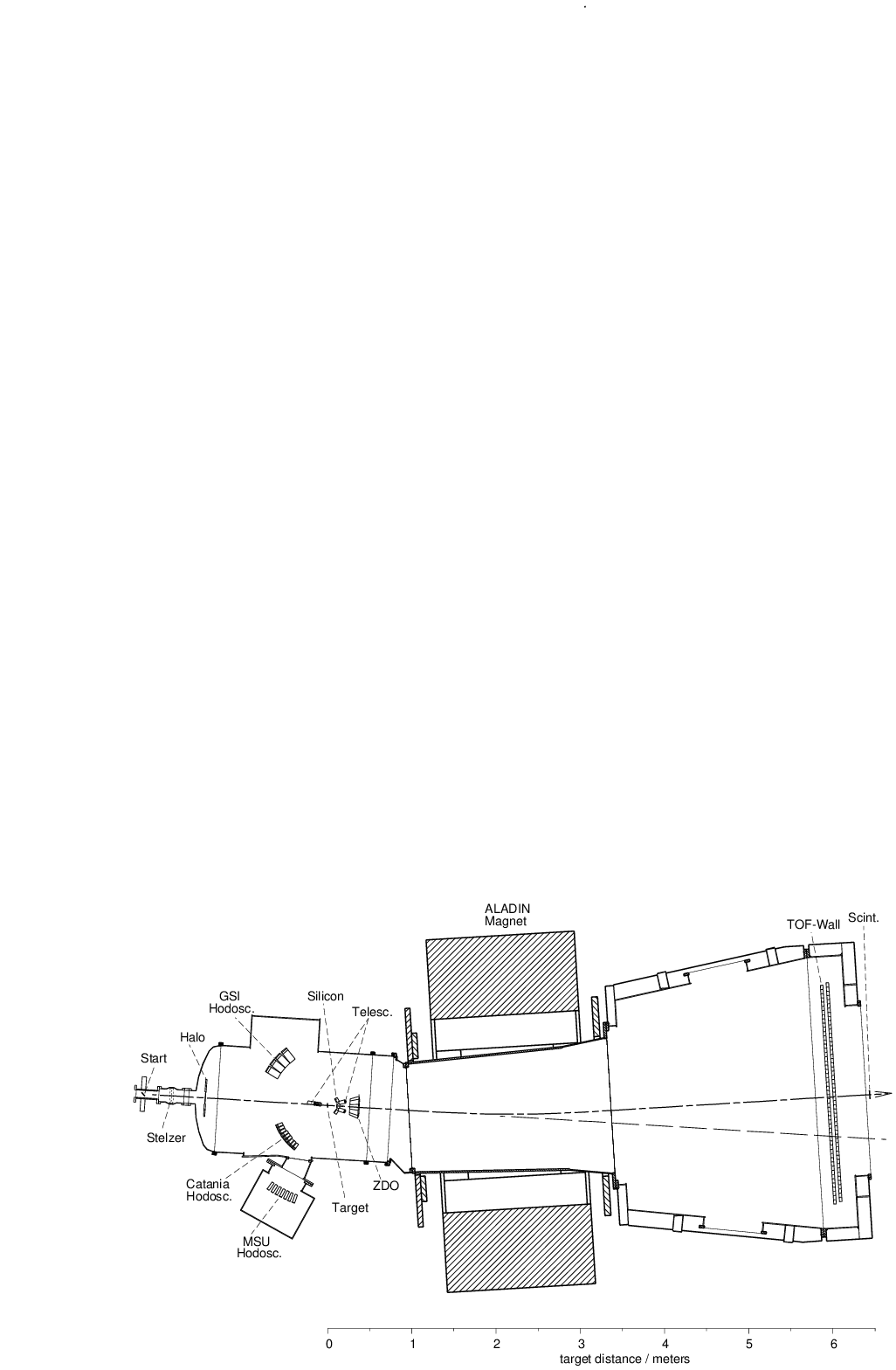}}
\caption{\sl Experimental setup}
\label{setup}
\vspace{0.1mm}
\end{center}

\end{figure}
\begin{figure}[b]
\begin{center}
\fbox{\epsffile[0 0 423 213]{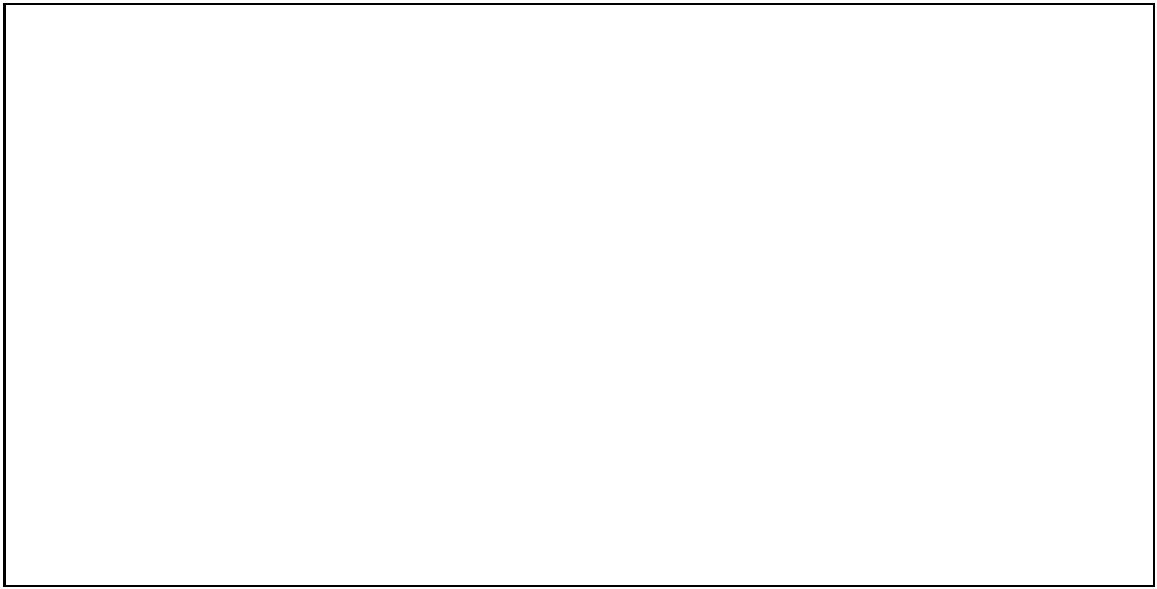}}
\caption[]{\sl $\Delta E$ vs $E$ (left), PID vs energy in MeV (middle) and
PID spectrum (right)}
\label{hodspec}
\vspace{0.1mm}
\end{center}
\end{figure}

The experiment was performed at the SIS accelerator at GSI, Darmstadt.
The experimental setup ist shown in Fig.~\ref{setup}.
A 25 mg/cm$^2$ Au target was irradiated by a 1 A$\cdot$GeV Au beam at a
beam intensity of $10^6$ particles per second. 
The forward hodoscope ZDO consisted of 36 phoswich detectors with a 
coverage of 6.5$<\theta<$21.5$^0$ and 0$<\phi<360^0$.
Three small--angle high--resolution $\Delta E-E$--hodoscopes with a total of
216 elements were placed at laboratory angles $\theta=100-150^0$.
Each of the 216 elements consisted of a silicon detector of 300 $\mu m$
thickness and a cesium iodide detector of 6 cm length for 160 detectors and 
10 cm length for 56 detectors, read out by photodiodes. Isotopes were resolved 
for $Z=1-4$. 
The energy calibration was performed according to \cite{twen90} and an energy 
resolution of about 1\% was achieved. From the cross 
calibration between detectors we got an additional smearing of 1\% resulting 
in an overall  uncertainty of $\approx2$\%. 
The superposition of the
$\Delta E$--$E$ distributions of all detectors is shown in Fig.~\ref{hodspec}. 
The separation of the Hydrogen isotopes (at $Z=1$), the Helium isotopes (at 
$Z=2$) and the Lithium isotopes (at $Z=3$) is clearly visible. \\[0.5ex]
The trigger required two valid hits in the ZDO and two valid hits in the 
combination of the three hodoscopes. For event characterisation we used
the ALADiN Time-of-flight wall, which is described in detail elsewhere 
\cite{sch96}. The variable $Z_{bound}$ in the TOF array which is the summed 
charge in forward angles excluding the $Z=1$ particles serves well as a 
criteria of the centrality of a collision \cite{sch96}. Analogous event
selection criteria as in \cite{poc95,sch96} were used. By the 
measurement of the decaying projectile spectator fragment charges with 
$Z\geq2$i and the neutrons (\cite{sch96}) an excitation energy $E^*$ and a 
mass of the prefragment $A_0$ could be assigned to every \Zbound~bin 
\cite{gross97}. 
The mass of the prefragment $A_0$ was used in the present analysis to extract 
the freeze out density from the radii derived by \pp~ and
\dalf~ correlations.\\[0.5ex]
\begin{figure}[htb]
\begin{center}
\fbox{\epsffile[0 0 423 283]{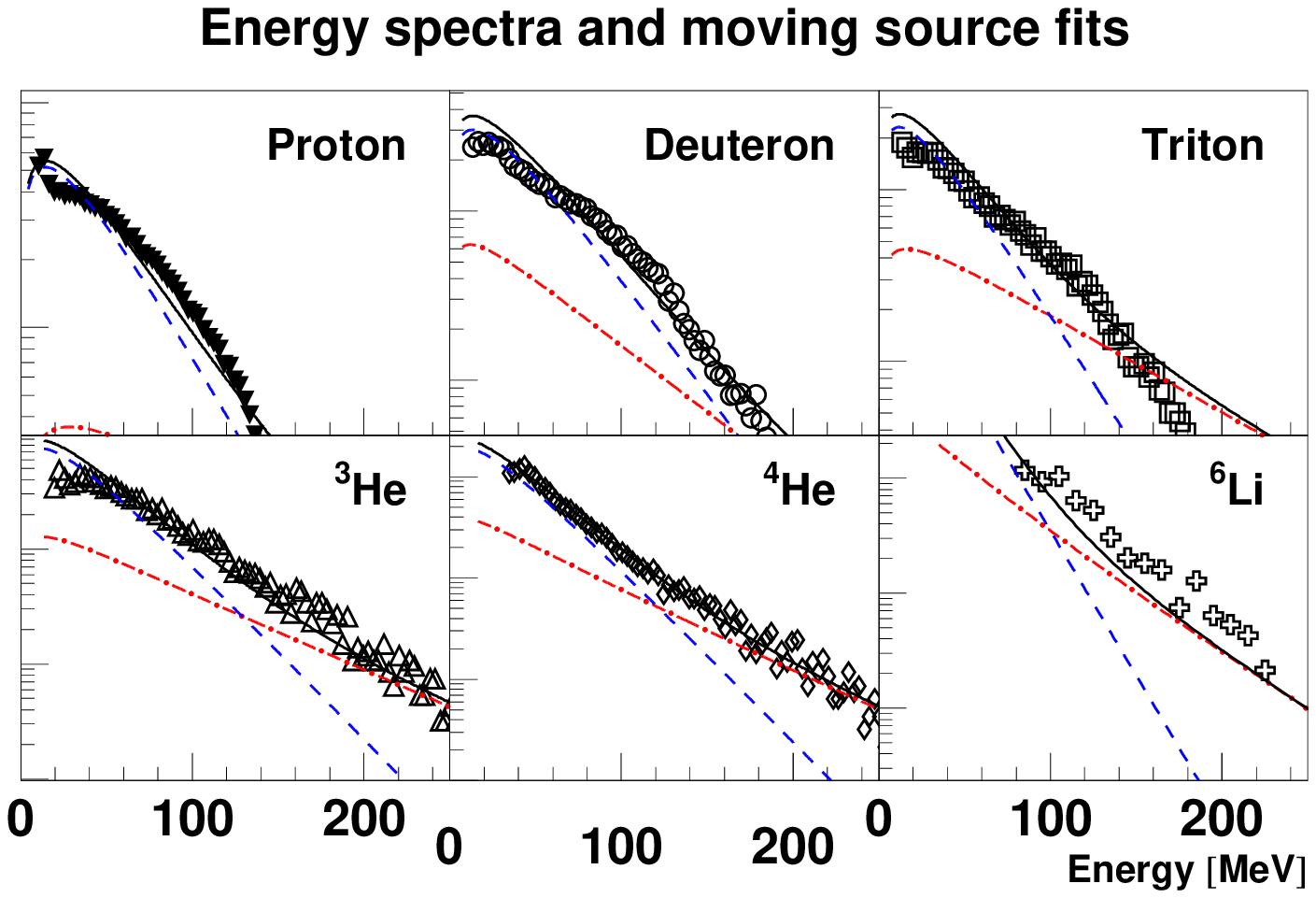}}
\caption[]{\sl Energy spectra  at $\langle\theta\rangle=35^0$
and 2 source fit for 1 A$\cdot$GeV\newline 
dashed lines $\equiv$ target source, 
dot dashed lines $\equiv$ midrapidity source}
\label{mvsource}
\vspace{0.1mm}
\end{center}
\end{figure}
In Fig.~\ref{mvsource} energy spectra for light particles together with
moving source fits assuming three sources in a symmetric system are presented. 
The moving source fits gives the following results:
\begin{ownenum}
\item The main contribution of the spectra can be described by a single source,
the target source moving with less than 3\% $c$ in the laboratory frame.
\item The inverse slopes of the spectra decrease with increasing mass of the 
particle species even for the most central collisions. 
\item The inverse slopes increase continuously with increasing centrality. 
This rise is partly due to the increase of bounce with decreasing impact 
parameter \cite{gross97}, partly to the increasing excitation energy with 
decreasing impact parameter. Its difference to temperatures extracted from 
isotope yields or excited states can be partly assigned to the effect of the 
Fermi motion \cite{bauer95} and to a significant part of pre--breakup emission 
of light particles \cite{trautmann97}.
\end{ownenum}
\begin{figure}[htb]
\begin{center}
\fbox{\epsffile[0 0 423 283]{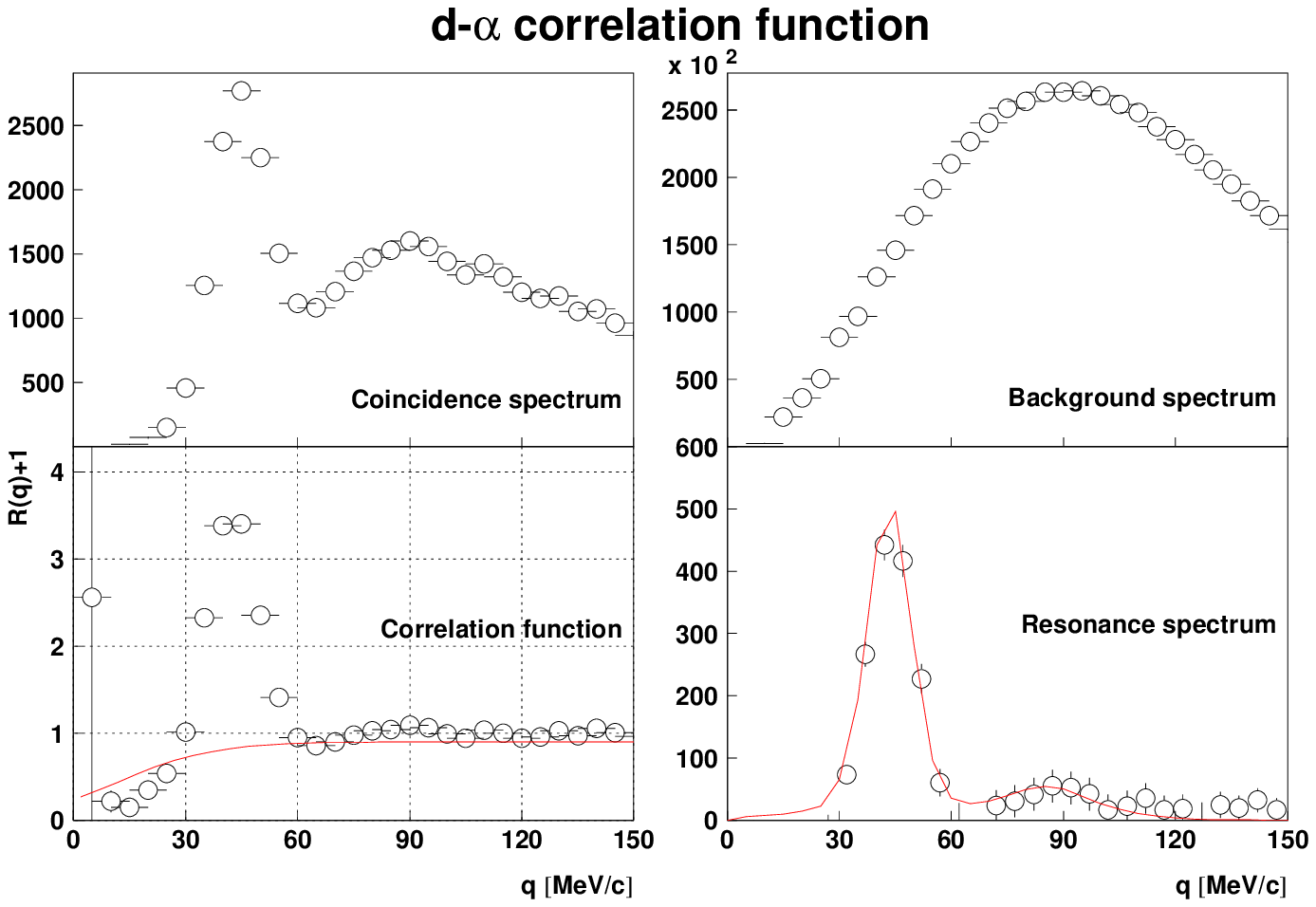}}
\caption[]{\sl Coincidence yield in the \dalf~correlation function
(upper left), uncoincident yield from the event mixing technique  
(upper right), ratio of coincident and uncoincident yield (symbols)
and assumed nonresonant part (line, lower left), extracted yield (symbols) 
compared with simulated yield (line, lower right)}
\label{npartplot}
\vspace{0.1mm}
\end{center}
\end{figure}
In order to construct the correlation function for a specific particle pair one
combines all particles of one species with all particles of the other species 
within one event and sorts the pairs according to their relative momenta. The 
coincidence spectrum shows the resonances of interest on top of a broad 
background distribution which is determined by phase space and acceptance of 
the hodoscopes. This 
background was determined with the event mixing technique \cite{kop74}, 
combining particles from different events. The correlation function is 
constructed by dividing the coincidence yield by the background yield and 
normalizing it to 1 for large relative momenta as shown in 
Fig.~\ref{npartplot} for the \dalf~correlation function.  For the 
determination of a
temperature one is interested in the absolute yields of the different states.
First, the nonresonant part of the correlation function has to be subtracted. 
This part comes from Coulomb repulsion and quantum mechanical effects
and was determined by scaling non-resonant correlation functions 
like \tritri~ or \trihed. After subtracting this non resonant part of the 
correlation function and multiplying the resulting spectrum with the integral 
coincidence yield a resonance spectrum is derived. This is
shown in the lower right panel of Fig.~\ref{npartplot}. 
The spectrum is compared to a Monte Carlo calculation for a certain 
emission temperature (here 5 MeV), which includes the 
first five excited states of \lis, the detection efficiency and resolution 
of the detector. Details of the Monte Carlo calculation are presented in
Ref.~\cite{serfling97}.
\section{Temperature extraction from excited states}\label{temp}
\begin{figure}[htb]
\begin{center}
\fbox{\epsffile[0 0 423 283]{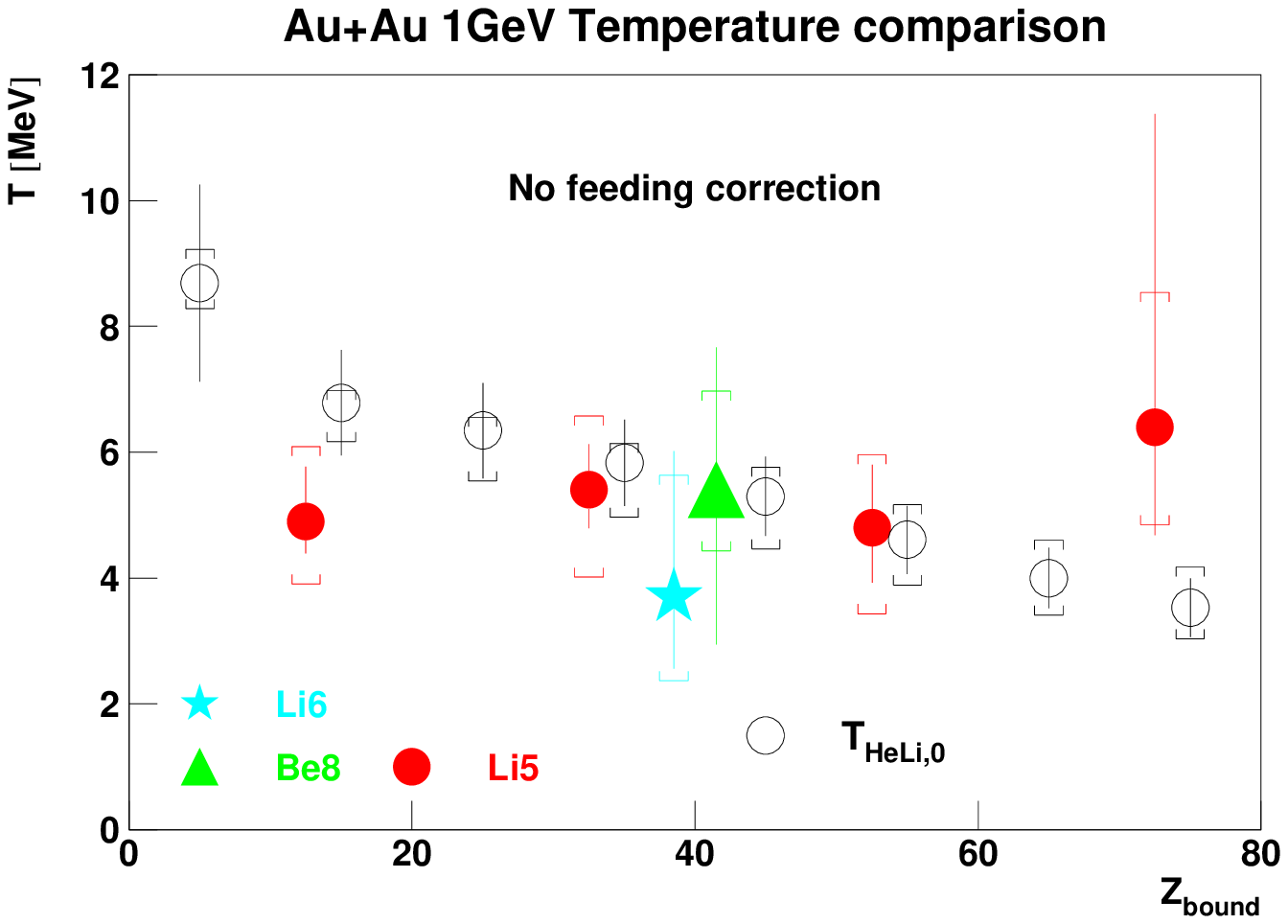}}
\caption[]{\sl Apparent temperatures from excited states for different 
thermometers vs \Zbound~and comparison with isotope thermometer \Theli~from 
the target spectator without feeding correction --- 
The error bars represent the statistical uncertainty, the
systematic uncertainty is indicated by the brackets}
\label{t_allplot}
\vspace{0.1mm}
\end{center}
\end{figure}

Details of the level schemes for the analyzed thermometers are presented in 
table~\ref{tabexctemp}
\begin{table}[htb]
\begin{center}
$
\begin{array}{||c|c|c|c|c|c||}\hline\hline
\mbox{Thermometer} & \mbox{Decay channel} & \mbox{Spin} J^{\pi} & 
\mbox{Energy [MeV]} & \Gamma \mbox{[kev]} & q \mbox{[MeV/c]} \\[1.5ex] \hline
     & p-\alpha      & 3/2^- & \mbox{G.S.} & 1500  &  54.3 \\
\raisebox{1.5ex}[-1.5ex]{\lif}     
     & d-^3\mbox{He} & 3/2^+ & 16.66       &  200  &  24.7 \\  \hline
                     
     & d-\alpha      & 3^+   &    2.17     &   24  &  41.5 \\
\lis & d-\alpha      & 2^+   &    4.31     & 1700  &  84.1 \\ 
     & d-\alpha      & 1^+   &    5.65     & 1500  &  99.1 \\  \hline
                     
     & \alpha-\alpha & 0^+   & \mbox{G.S.} &6.8~eV &  18.3 \\
\bea & \alpha-\alpha & 2^+   &    3.04     & 1500  & 108   \\
     & p-^7\mbox{Li} & 1^+   &   17.64     &   10.7&  27   \\  \hline \hline
                     
\end{array}
$
\end{center}
\caption[]{\label{tabexctemp} Level schemes of analyzed particle--unstable states
(from \protect\cite{ajz84})}
\end{table}
\begin{figure}[htb]
\begin{center}
\fbox{\epsffile[0 0 423 423]{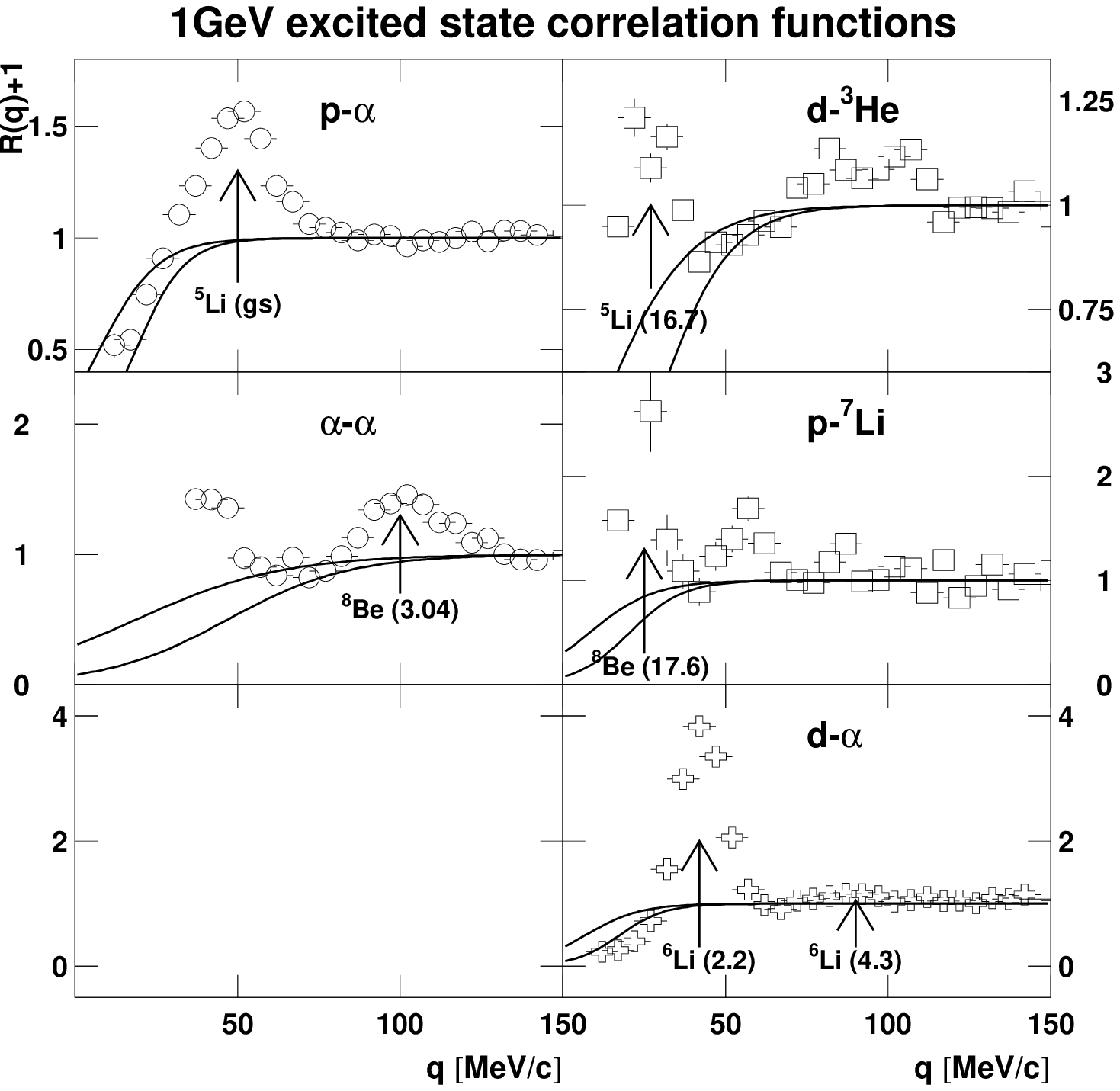}}
\caption[]{\sl {\bf First row:} Ground state (\palf, left) and first excited 
state (\dhed, right) of~\lif \newline
{\bf Second row:} Ground and first excited state (\alfalf, left), 
fourth to sixth  excited state (\plisi, right) of~\bea\newline
{\bf Third row:} 
First, third and fifth  excited state of (\dalf, right) of~\lis}
\label{li5plot}
\vspace{0.1mm}
\end{center}
\end{figure}
For the \lif ~ thermometer the ratio of the first excited state at 
$E^*=$16.66~MeV to the ground state was used. The wide ground state was fitted 
with the R--matrix formalism with values taken from \cite{ajz84}. 
The two correlation functions together with a 
minimum and a maximum background are shown in the
first row of Fig.~\ref{li5plot}. The well known peak of 
$^9$B$\rightarrow p+^8$Be$_{g.s}\rightarrow p+(\alpha+\alpha)$ just 
above threshold could not be resolved in the 
experimental correlation function. The apparent temperature for \lif~ and
the other excited state thermometers is determined by fitting the yield ratios
of simulated resonance spectra for a range of input temperatures to the 
experimental yield ratios and using to the exponential temperature formula
\begin{equation}\label{tempform}
\frac{Y_2}{Y_1}=\frac{2J_2+1}{2J_1+1}\cdot\exp\left(-\frac{\Delta E}{T}\right)
\end{equation}
\bea~ decays in the ground and the first excited state into \alfalf, 
the fourth to sixth excited states decay into $p+^7$Li (Fig.~\ref{li5plot} 
second row). The ground state ($q=18$~MeV/c) is too close to the detection 
threshhold of $q=15~MeV/c$, so that we used the ratio of 
the fourth excited state at $E^*=17.64$~MeV and the first excited state at
$E^*=3.04$~MeV for the temperature 
determination. The disadvantage of this thermometer is a rather small yield
of \bea~resulting in large statistical uncertainties.
Thus the apparent \bea--temperature was  determined without any centrality 
selection criterium.\\[1.5ex]
For the \lis~ thermometer we have chosen the ratio of the 4.31~MeV and
5.65~MeV to the 2.17~MeV state. All three states decay into 
\dalf~(Fig.~\ref{li5plot} lower right panel). Both resonances of \lis~ were 
described with a Breit--Wigner function (see Fig.~\ref{npartplot}).
Because of the small energy difference strong feeding distortions of
the \lis~population are expected.  Additionally a small change in the the
background correlation function results in a big change of the relative yields 
due to the large values of the background yields resulting in large systematic 
errors. No centrality selection criterium was applied on the 
\lis~ thermometer.\\[1.5ex]
The \lif~ thermometer yields a rather constant apparent temperature of 
$T\approx5$~MeV independent of impact parameter (Fig.~\ref{t_allplot}).
The apparent temperature values are in the same range as 
reported elsewhere \cite{poc85,poc87,kun91,nay92,schwarz93}.\\[0.5ex]
On the other hand, the extracted apparent temperature does not show a 
systematic rise as does the temperature \Theli, which was derived from the 
relative isotope ratios of Helium and Lithium \cite{trautmann97} 
(open circles in Fig.~\ref{t_allplot}).\\[0.5ex]
For \lif~ the dependence on the kinetic energy of the particle pair
was also investigated but no systematic trend could be established.
\section{Freeze out density from \pp~and \dalf~correlations}\label{density}
\begin{figure}[p]
\begin{center}
\fbox{\epsffile[80 160 505 690]{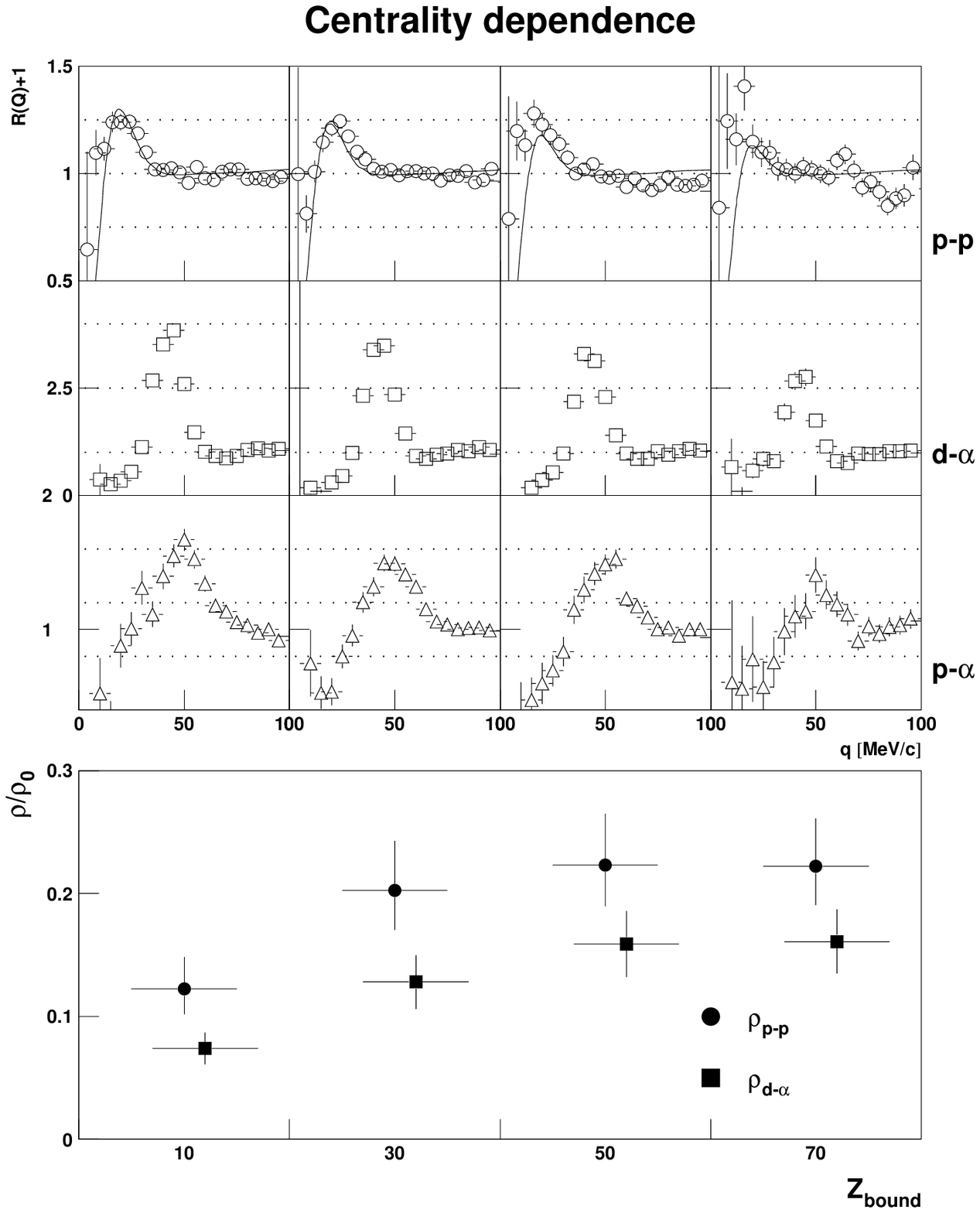}}
\caption[]{\sl {\bf First row:} \pp~correlation function for four different cuts
in \Zbound~(see x axis on fourth row) and comparison to simulated correlation 
functions for $r_0=8.0-9.5$~fm respectively\newline
{\bf Second row:} \dalf~correlation function for the same cuts as \pp~in
\Zbound\newline
{\bf Third row:} \palf~correlation function for the same cuts as \pp~in
\Zbound\newline
{\bf Fourth row:} Mean density from extracted radii and prefragment mass $A_0$
vs \Zbound}
\label{ppcorr}
\vspace{0.1mm}
\end{center}
\end{figure}

In the first row of Fig.~\ref{ppcorr} the \pp~ correlation function for four
different cuts in \Zbound~ together with a simulation from the 
Koonin--Pratt--formalism 
\cite{koonin77} for $r_0=8.0,8.5,9.0,9.5$~fm and $\tau=0$~fm/c is 
shown.\\[0.5ex]
The peak height of \dalf~ and \palf~ correlations should also be 
sensitive on the freeze out radius \cite{boal86,boal90}. The dependence of
those on \Zbound~ is shown in the second and third row of Fig.~\ref{ppcorr}. 
For the \dalf~ correlations the integral of the first and second maximum was 
compared to the integral of theoretical \dalf~ correlation functions 
\cite{poc87,boal86,boal90} in order to get absolute numbers for the \dalf~ 
freeze out radius. The hard sphere radii vary from 10~fm for \Zbound$<20$ to 
11~fm for \Zbound$>60$.\\[0.5ex]
Taking into account the increasing mass of the target spectator $A_0$ with 
increasing impact parameter we determined a mean freeze out density. This mean
freeze out density varies from $0.2\rho_0$ ($0.17\rho_0$) in peripheral to 
$0.1\rho_0$ ($0.07\rho_0$) in central collisions for the \pp~(\dalf) 
correlations (fourth row of Fig.~\ref{ppcorr}). These densities are smaller 
than the standard density of $0.3\rho_0$  which is usually  used in theoretical 
calculations. A freeze out density which decreases with decreasing impact 
parameter was already proposed by \cite{warren96} from INC 
calculations.\\[0.5ex]
\begin{figure}[hp]
\begin{center}
\fbox{\epsffile[80 160 505 690]{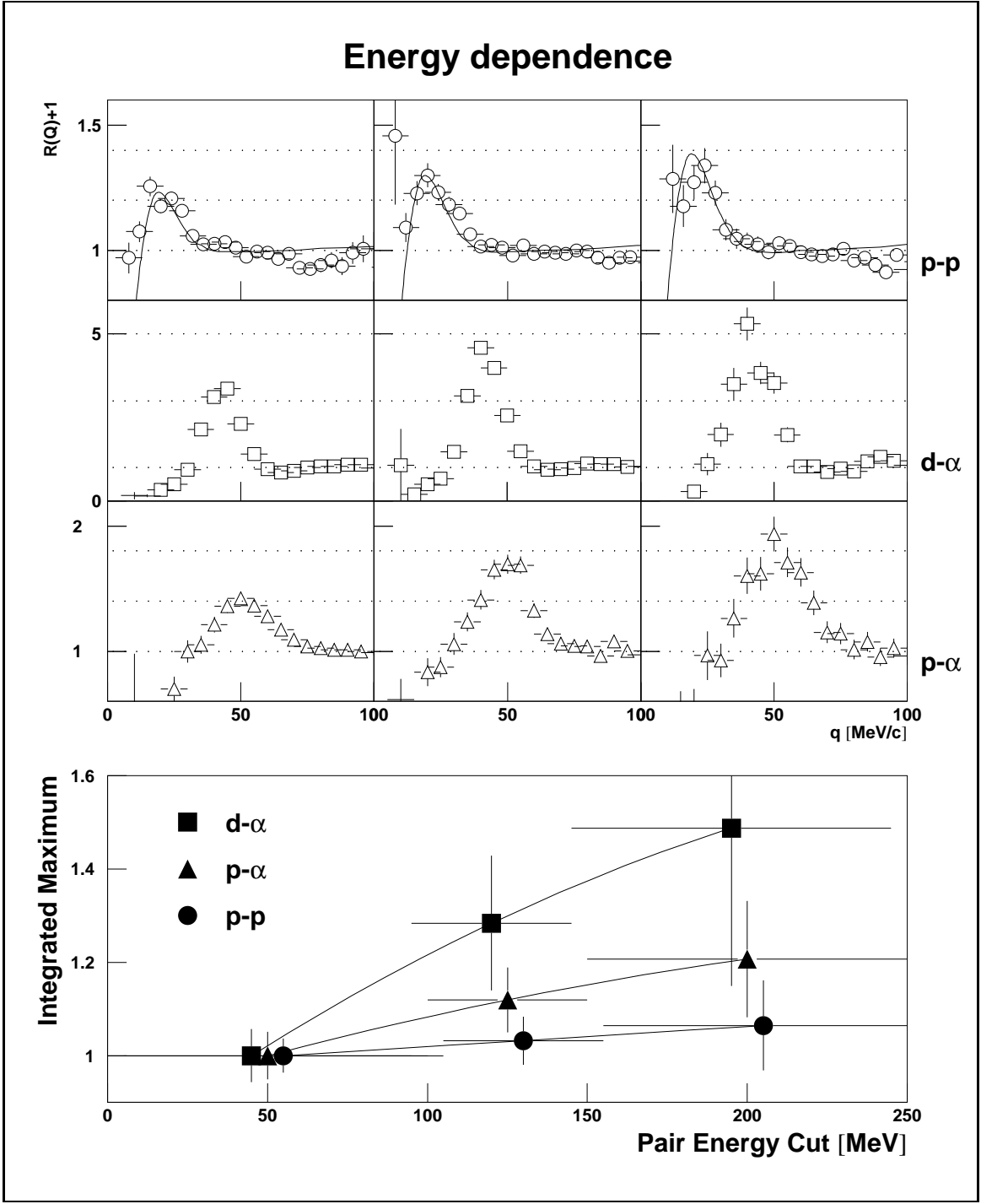}}
\caption[]{\sl {\bf First row:} \pp~correlation function for three different 
cuts in summed energy (see horizontal error bars in bottom panel) and 
comparison to simulated correlation functions for 
$r_0=9.0$~fm, $r_0=8.5$~fm and $r_0=7.5$~fm respectively\newline
{\bf Second row:} \dalf~correlation function for the same cuts as \pp~in
summed energy\newline
{\bf Third row:} \palf~correlation function for the same cuts as \pp~in
summed energy\newline
{\bf Fourth row:} Rise of the integrated maximum in the correlation function 
with the summed energy for \pp, \dalf~ and \palf~correlations}
\label{expansion}
\vspace{0.1mm}
\end{center}
\end{figure}
In Fig.~\ref{expansion}  \pp, \dalf~ and \palf~ correlations are shown
for for three different ranges of summed energy. The peak heights of the 
correlation functions and as a consequence the 
extracted radii vary strongly with the summed energy of the particle pairs
(Fig.~\ref{expansion}). Faster particles seem to be emitted from a smaller
source. This may be interpreted as a time ordered emission during expansion.
\section{Comparison to Feeding calculations}\label{feeding}
\begin{figure}[htb]
\begin{center}
\fbox{\epsffile[0 0 423 283]{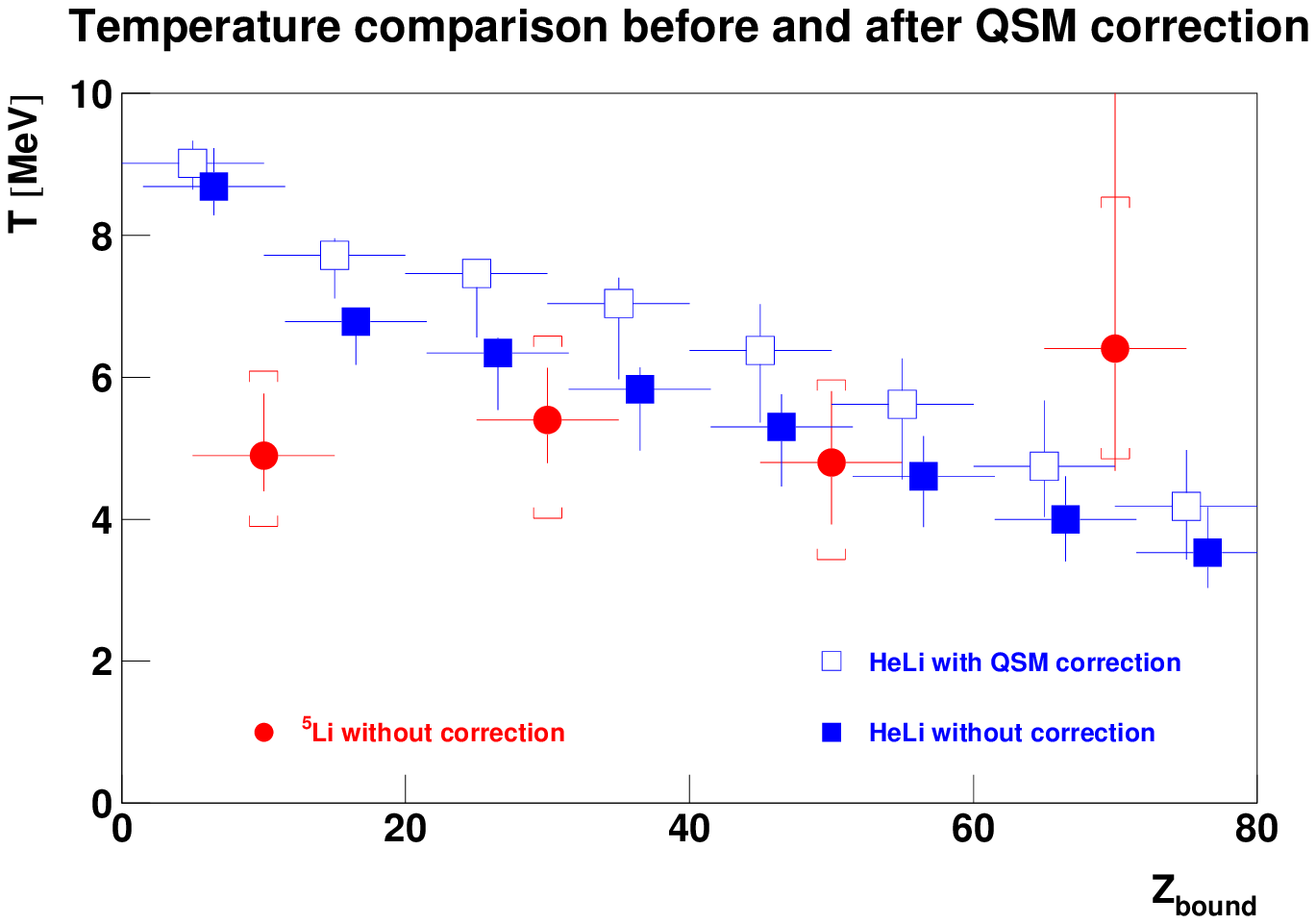}}
\caption[]{\sl \Theli~ before and after density dependent feeding correction 
according to QSM and comparison to apparent \lif~temperature 
}
\label{t_qsm}
\vspace{0.1mm}
\end{center}
\end{figure}
For the analysis of feeding distortions we used the QSM code \cite{hah88}.
The {\bf Q}uantum {\bf S}tatistical {\bf M}odel is based on 
the assumption of thermal and chemical equilibrium. This model incorporates 
infinite system size. The ground and excited states of nuclear
states up to $Z=13$ are included according to published values \cite{ajz84}. 
No continuous states are included in the QSM. The excluded volume of a 
fragment is the sum of the nucleon volumes ($^3$H and $^3$He have the same 
excluded volume). The states are 
populated according to one global temperature $T$, density $\rho$ and $N/Z$ 
ratio. The initial particle ratios are modified due to the sequential decay of 
particle unbound states (feeding) and one
has to distinguish between an initial and a final distribution. The final ratio
was used to calculate yields of ground and excited particle
unbound states. By applying the exponential temperature formula 
(equation~\ref{tempform}) 
we derived a relationship between the initial temperature 
and the apparent \lif, \bea~ and \lis~ temperature respectively. 
The comparison of the experimental yield with the QSM calculation gives the
following results:
\begin{ownenum}
\item The constant correction factor for the \Theli~ thermometer of 1.2 
used by \cite{poc95} was determined assuming a constant freeze out 
density $\rho=\rho_0/3$. However, the extracted freeze out radii from 
\dalf~correlations measured in the same experiment can best be described 
with a density 
varying from $0.17\rho_0$ for the most peripheral to $0.07\rho_0$ for
the most central collisions. Because of the density dependence of the QSM
correction the correction factor for the isotope thermometer becomes 
centrality dependent ranging from 1.2 for the most peripheral to almost 1 for
the most central collisions. This reduces the observed 
effect of a steep rise in temperature for the most central collisions. 
The \Theli~thermometer before and after QSM correction is shown in 
Fig.~\ref{t_qsm}.
\item According to QSM there is no feeding correction factor needed for the
extraction of the  \lif~temperature. Thus within QSM and the assumption of a 
global freeze out for all particle species only part of the 
discrepancy between excited state and isotope temperatures can be explained. 
\item Additional feeding calculations for the \lif~thermometer were performed 
analogous to \cite{xi96} (see also \cite{tsang97}). The correction for 
the \lif~ thermometer from this code is less than 1 MeV.
\end{ownenum}
\section{Summary}\label{sum}
The extraction of apparent temperatures from of excited state population
ratios for the system Au+Au at 1~A$\cdot$GeV was presented and temperature 
values of $T\approx5$~MeV are derived. In contrast
to the isotope thermometer presented in \cite{poc95,trautmann97} 
the apparent temperature stays constant with decreasing impact parameter thus 
increasing excitation energy. The freeze out densities extracted by the 
means of \dalf~ correlations and used as input for QSM calculations seem to 
favour a decreasing freeze out density with increasing excitation energy. 
After feeding correction one gets a rather slow rise in emission temperature 
starting at $T=5$ MeV for the most peripheral and reaching around $T=9$ MeV for
the most central collisions with the isotope thermometer while the excited 
state thermometer saturates around $T\approx6$ MeV emission temperature. 
This discrepancy is even more pronounced in Ref.~\cite{schwarz97} and explained
by different freeze out times for isotope and excited state thermometers.

\end{document}